The Hardcore Brokers:

Core-Periphery Structure and Political Representation in Denmark's Corporate Elite Network

Lasse F. Henriksen, Jacob Lunding, Christoph H. Ellersgaard and Anton G. Larsen

Abstract

Who represents the corporate elite in democratic governance? In his seminal work on the "corporate inner circle", Useem (1986) identifies the emergence of three network-related mechanisms that shaped the composition and political organization of American and British corporate elites in the postwar era: organizational brokerage, elite-level social cohesion and network centrality. Subsequent research has found similar dynamics at play across a variety of capitalist societies but all studies on corporate political representation rest on network analyses of a highly select sample of leaders from the top ranks of very large publicly listed firms. We cast a wider net. Analyzing new population data on all members of corporate boards in the Danish economy (~200,000 directors in ~120,000 boards), we locate ~1,500 directors that operate as brokers between local corporate networks and measure their network coreness using k-core detection. We find a highly connected network core of ~275 directors, half of which are affiliated with smaller firms or subsidiaries. Statistical analyses show a strong positive association between director coreness and the likelihood of joining one of the 650 government committees epitomizing Denmark's social-corporatist model of governance (net of firm and director characteristics). The political network premium is largest for directors of smaller firms or subsidiaries, indicating that network coreness is a key driver of business political representation, especially for directors without claims to market power or weight in formal interest organizations.

INTRODUCTION

Who represents the corporate elite in democratic societies? The representation of corporate interests in democratic governance is a key concern in sociology and political economy (Mills and Domhoff 2023; Comet 2019; Useem 1986; Domhoff 1967; Mills 1956; Hilferding 1910; Marx 1859). A large empirical literature documents the central role of corporate elite networks in shaping the political orientation of businesspeople and their involvement in democratic governance: how cross-firm elite interactions can modify the political orientation of corporate executives (organizational brokerage), how social network cohesion shapes the collective action potential of managerial elites (cohesion), how movement into central network positions can heighten their elite status (centrality), and how these elite network characteristics together enhance the corporate elite's access to government policymaking (Mills and Domhoff 2023; Mizruchi and Hyman 2014; Burris 2010; Mizruchi 2013; Useem 1986; Mokken and Stokman 1978).

To date the empirical scope of this literature is limited to considering the elite network of an economy's largest corporations. We combine a new administrative register on the members (~200,000) of all corporate boards (~120,000) in the Danish economy with a hand-collected data set on government committee members (597 out of 650 committees contain at least one director, and 3000 directors are committee members). This data allows us to analyze the association between the position of directors in a complete elite population network and their involvement in democratic governance. We posit that elite network centrality is best conceptualized in terms of core-periphery structure and propose k-core detection as method for measuring a director's coreness. We also suggest that sorting in the core-periphery structure is contingent on a director's local brokerage opportunities. We then analyze the association between these contingent network characteristics and likelihood of government committee memberships. We find that local brokerage and high network coreness are strong predictors of a director's likelihood of being on a committee net of director and firm characteristics. We also find that high network coreness is more positively associated with committee memberships for directors in smaller independent corporations and subsidiaries (of which there is a surprisingly large number) than directors of large corporations. These findings suggest that research on corporate elite networks should include directors from a wider distribution of corporations in future studies.

THE CORPORATE ELITE NETWORK AND POLITICAL REPRESENTATION

Michael Useem's (1986) work on the "inner circle" network has been especially important in shaping thinking about corporate elite representation in democratic governance. He traced the historical emergence of a cohesive network made up of board connections among the leaders of the largest American and British corporations. Useem's studies brought attention to the role played by elite social connections in shaping the worldview of corporate leaders and as a pathway for them into the political sphere (Useem 1986, 1978).

The inner circle network and its implications for elite collective action has been studied intensely across a broad variety of socioeconomic contexts (Mills and Domhoff 2023; Hong, Lee and Yoo 2021; Benton 2019; Comet 2019; Heerwig and Murray 2019; Chu and Davis 2016; Mizruchi 2013; Chung 2003). While more recent research from the US (Chu and Davis 2016; Mizruchi 2016; Benton 2019), the Netherlands (Heemskerk and Fennema 2009) and Switzerland (AUTHOR(S)) indicate tendencies of elite fracturing in the corporate network, studies from the European Union overall (Heemskerk, Daolio and Tomassino 2013) and from France (Comet 2019) and Denmark (AUTHOR(S)) suggest the continued existence of densely connected inner circle networks that remains a central pathway for corporate leaders into democratic governance.

All the above research follows Useem's original recipe for identifying the boundaries and network structure of the "corporate inner circle": A sample of the largest corporations in the economy identifies a set of *ex ante* elite corporate actors and their connections are then traced by mapping the network of directors that emerge from their multiple board appointments (what is also termed the interlocking directorate). Useem himself recognized the limitations of looking solely on the largest corporations: "Since the degree of control may range from influence over a few companies to a voice in the policies of many, the boundary between the inner group and the remainder of the capitalist class is diffuse rather than sharp" (Useem 1978: 227). Even before then Allen argued that while "it is possible to construct a saturation sample of a population which has been delimited in accordance with relevant theoretical criteria […] [t]he most satisfactory sampling design for structural analysis is a saturation sample of the entire universe or population" (1974:396).

Despite these early cautions, most contemporary research into corporate elite networks present no explicit or convincing theoretical argument for the continuation of the empirical focus on large corporations. From a network analytic perspective this is problematic because the structural properties of networks essentially reflect the selection of actors and relations for analysis (Kossinets 2006; Laumann, Marsden, and Prensky 1989). Barnes (2017) for example points to key network locations beyond the largest corporations, including the boards of non-profits, where business leaders coordinate elite political action. In a recent study of 3500 Canadian boards, Huijzer and Heemskerk (2021) also found that firm-based sample restrictions strongly affect the observed structural properties of the sampled network, suggesting that data completeness is essential to reach accurate estimates of structure in inter-firm networks.

The choice of focusing on the leaders of large corporations was meaningful for Useem in his analyses of the 1970s inner circle network in the sense that he saw corporate concentration and the rise of centralized managerial power in firms as key contingencies allowing the corporate elite to emerge as a collective actor in the first place. While the concentration of corporate control has accelerated in most capitalist countries since the 1970's, which would justify the sustained focus on large firms in the contemporary literature, the legal-organizational structure of these firms is also undergoing significant change. Organizational analysts have for long stressed the increased "blurring", or "fissuring", of firm boundaries associated with a host of restructuring processes that is reconfiguring the power structure of corporations (Davis 2022), including for example the vertical disintegration (Whitford and Zirpoli 2016), organizational fissuring (Weil 2014), internationalization (Murray 2017), and financialization (Seabrooke and Wigan 2017; Davis and Kim 2015; Davis and Mizruchi 1999) of the modern firm. At the same time institutional complexities challenges the prerogative of centralized managerial power in corporations and increasingly result in a diversification of managerial power (Svejenova and Alvarez 2022, 2015) and mounting shareholder pressures (Benton 2019, 2016; see also Davis 1991) that may also affect the structure of the corporate elite networks in important ways. The implications of such organizational and managerial reconfigurations for the structure of corporate elite networks are understudied.

As a result, research on corporate networks may benefit from a more exploratory stance on what actors and relations to include in analyses of corporate elites, and from pursuing more open-ended investigations of who represents corporate interests in democratic governance. The increased availability of large-scale corporate governance databases and registries that extends beyond the social connections of large-firm directors presents an opportunity to gain insights about how corporate networks in the wider business community are structured, and whether the structure of these expansive networks can researchers better understand who represents corporate interests in democratic governance and what mechanisms drive the representation. Recent political economy scholarship on the structure of global corporate networks have made good use of more inclusive board and ownership data (AUTHOR(S)), for example with analyses of asset concentration (Fichtner, Heemskerk and Garcia-Bernardo 2017; Vitali, Glattfelder and Battiston 2011) and transnational elite community structures and centralization (Takes and Heemskerk 2016), but we are still to see how such data may inform studies of political representation among national corporate elites. In what follows we develop a new approach to identifying the political representatives in the corporate elite network based on the universe of companies in a national economy.

THE NETWORK ELITE OF CORPORATE DIRECTORS: THE CORE OF BROKERS

Our key aim is to explore the extend to which the political actors in the corporate elite can be inferred entirely from the network structure of the business community, and how much their political status depends on the network structure vs. the characteristics of the directors and companies in the network. We propose an approach that identifies the corporate elite from the core-periphery structure of the country's entire corporate network. The core-periphery structure, the organization of a network into a well-connected community of core actors and a lesser-connected periphery, is a powerful image describing many large-scale social, biological and technological networks (Alba and Moore 1978; Borgatti and Everett 2000; Gallagher, Young, and Welles 2021; Laumann and Pappi 2013; Malvestio, Cardillo, and Masuda 2020). The term network core centers on the two main network-positional characteristics already stressed in corporate elite research as predictors of elite integration and political representation: cohesion and centrality (Borgatti and Everett 2000; Batagelj and Zaversnik 2003).

Consider a weighted undirected network of corporate directors where $i$ is connected to $j$ with weight $==1$ if they serve on at least one board together at time $t$, and where $i$ is connected to *j* with weight = 0.5 if they share a common connection at a board but do not themselves serve on the same board. Apply on this network a weighted *k*-core decomposition that identifies the nested structure of successively more and more connected-and-cohesive elite groups in the largest network component at each time point *t* (see also, AUTHOR(S)). The *k*-core decomposition prunes a network by sequentially removing nodes with a minimum degree *k* (the sum of weights in *i*'s ego network) up until the point of network degeneracy, that is, when the network disintegrates entirely. The innermost *k*-core of the network is characterized by having the degeneracy *k*-minimum weight, which is the maximum *k*-core in the network. The *k*-core score of nodes is otherwise known as coreness.

One elegant property of the *k*-core decomposition is that it identifies the discrete inner core of the network (to the extent that the network is structured along a single core-periphery axis), which reveal a strongly integrated cluster where all directors are connected to at least *k* directors in the cluster (Heemskerk, Daolio, and Tomassino 2013; Huijzer and Heemskerk 2021). Maximal *k*-cores are highly resilient network structures that effectively diffuse information and norms, propel collective action and are robust to external chocks (Al-garadi et al. 2017; Kong et al. 2019; Malliaros et al. 2020). At the same time, the approach recognizes that core-periphery structures are in the last instance nested community or authority structures (Benton 2016), with actors occupying positions closer to or further from the innermost core of the network (Gallagher et al. 2021). While membership of the maximal *k*-core of a network will indicate extraordinarily high connectivity-and-cohesion, the relative proximity to this centre indicates important variation in actors' relative connectivity-and-cohesion (Csermely et al. 2013). If the boundaries around elites are fuzzy, coreness will indicate an actor's relative position in an elite status hierarchy.

The *k*-core approach differs from most approaches to locating elite communities in a network, which most frequently consider degree centrality distributions. In work on corporate elite networks in general, and on the political business elite in particular, research has highlighted the influence of super-connectors that command a disproportionately high span of corporate control (Murray 2017; Chu and Davis 2016; Heemskerk et al. 2016; Useem 1986). Notions of super-connectivity are associated with Matthew effects, power- and Pareto law models (Barabási and Albert 1999), which focus on accumulative advantage mechanisms among the most central nodes. While degree centrality indicates the amount of direct exposure to other directors and firms, super-connectors can principally derive their status from a wide set of directors that are not mutually connected and therefore not necessarily embedded in a cohesive elite community characterized by network closure, or cohesion.

Other studies discuss the community structure of corporate elites, focusing mainly on network cohesion within network regions. Benton (Benton 2016; see also Benton and Cobb 2019) deploys a structural cohesion lens to get at the nested authority structure of US-based inter-firm networks, borrowing from Moody and White (2003) who defines structural cohesion as "the extent that multiple independent relational paths among all pairs of members hold it together" (2003:107). Structural cohesion highlights a network block's ability to resist disintegration or being "pulled apart by the removal of a subset of members" (White and Harary 2001). Cohesive network blocks make up tightly knit structures of overlapping cliques, a structural feature associated with robust elite collective action (Benton 2016; Benton and Cobb 2019; Heemskerk, Daolio and Tomassiono 2013). Where structural cohesion algorithms locate local cohesive blocks, the *k*-core decomposition identifies the global core-periphery structure, with actors high in coreness characterized by high global nested authority.

However, studies show that the accuracy of the *k*-core method in determining actor coreness within networks is sensitive to the presence of core-like clusters with a high *k*-core index but with very little cross-cluster connectivity or "spreading efficiency" (Liu et al 2015).

In our case, when highly overlapping clusters of large boards become sufficiently large, directors with many redundant connections within them begin to dominate the overall degree distribution of the network even if their connectivity is exclusively local. Since the overall *k*-core distribution of a network is partially tied to the network's degree distribution, local peaks in the degree distribution possibly distorts identification of the overall core-periphery structure. The logic here being that for actors to be considered valuable network agents beyond their local clique, they should provide unique connectivity to more distant actors, connecting parts of the network that would otherwise not be connected (Doreian and Woodard 1994; Freeman, Borgatti, and White 1991).

We adapt Liu's (2015) measure of diffusion importance based on local betweenness to identify what we term local brokerage. More specifically, we define the local brokerage for director *i* in an undirected network:

$$Local\ brokerage_i = \frac{local\ betweenness_i}{degree_i}$$

where the local betweenness of *i* is the sum of paths *i – j – h* where *i* is not directly connected to *h* and where the degree *i* is the sum of direct connections of *i*. We apply a brokerage threshold of 1 iteratively, thus filtering out directors with a value smaller than 1.

Apart from improving the accuracy of the *k*-core index this method has the added benefit of extending Useem's (1986) of "linkers" (multi-positionality directors) with a network structural feature commonly associated with organizational elites, namely brokerage power (Burt 1992, 2010). Because scholars from this elite theoretical perspective tend to work with relatively small sample sizes containing only large and relatively discrete corporate entities, multi-positionality constitute a significant network signal. In our highly granular company-level board data, which contains a substantial number of large and highly overlapping clusters[1], multi-positional directors do not necessarily contribute to overall elite social cohesion. The brokerage measure helps locate multi-positionals who are effectively cross-cluster brokers in the network.

CORE-PERIPHERY ANALYSIS

*Network sample and data sources*

We use data on boards of directors from Statistics Denmark's corporate governance registry (*Deltagerregisteret*) which contains time-stamped board of director data for all companies legally required to have a board.[2] The data contains descriptions of board roles ("Executive", "Chair" and "Ordinary Member") linked to unique company identifiers ("cvrnr"). We construct monthly weighted adjacency matrices according to the network definition laid out above.

We begin from an exhaustive sample of 157,521 unique boards with more than one serving director in legally registered Danish companies from 2010 through 2015. This period is chosen because this is when our data on corporate directors' political representation was collected (more on this data below). The total sample contains 222,783 unique directors. We restrict this sample to include only boards of public and private businesses classified as operating in a market for profit, amounting to 117,564 unique boards, 199,690 unique directors and 436,405 unique director-board positions across the period, an average of about 2.2 board positions per director.[3] This is our full analytic sample. To deal with seasonality in board turnover, our units of observation are board positions by month, where a position is active in a month if the director serves for at least one day during that month.

---

[1] Sometimes these large overlapping clusters express strongly connected firms within corporate groups and sometimes they express business-group like structures but with no mutual legal ownership.

[2] Independent business companies (defined in Danish legislation as "single-owner" companies or "stakeholder companies" (companies with personal ownership shared between two owners) are not legally required to set up a board of directors, and very rarely do so. While limited liability companies are also not legally required to set up a board of directors, 99.9% of these companies do according to our data.

[3] We recognize that there may be a host of other sources of social cohesion to the corporate elite. Domhoff's (1967) showed that the corporate elite not only forges internal connections through board interlocks but also through shared memberships of non-profit organizations, and social and cultural institutions. Barnes (2017) demonstrates that such shared memberships continue to be an important source of social cohesion in the American corporate elite. We restrict our analyses to firms operating for a profit because we wish to highlight

*Results from the core-periphery analysis*

Figure 1 plots the fraction of the largest component remaining after each step of the local brokerage pruning. The first step excludes 77 percent locally redundant directors in the largest component and the subsequent steps additionally exclude 20 percent. On average across the months 1550 local brokers remain after our filtering (range 1169 - 1892). The overwhelming majority of company directors therefore have very limited brokerage capacity. The remaining sample of local brokers enters the *k*-core decomposition.

< Insert Figure 1 about here >

As described above, we apply the *k*-core algorithm to the monthly adjacency matrices where directors are connected to their first (weight==1) and second neighborhood (weight==0.5). A director with 5 first neighborhood connections and 10 second neighborhood connections then has reach degree $k=10$.

Figure 2 report results from the *k*-core decomposition. Figure 2.A (panels 1-4) illustrates the nested structure of *k*-cores at selected *k*-steps and clearly illustrate the core-periphery structure for the month of October 2015. Figure 2.A1 shows the full network where all brokers have at least *k* 7.5. Panels 2.A2 and 2.A3 emphasize the k-cores at thresholds $k=20$ and $k=23$ respectively. Panel 2.A4 displays the nodes with maximal k-core score $k=26$, which is the degeneracy=26.5-0.5. In Figure 2.B1 we plot the fraction of brokers remaining at each step of the *k*-core decomposition for all months in the data. Panel 2.B2 shows that the

---

the potential role of smaller business firms in generating elite social cohesion and to make a succinct contribution to the board interlock literature.

average maximal coreness across all the months is 23 (range 20.5 - 27.5) and the average number of directors with maximal coreness is 278 (range 164 - 393).[4]

Since the coreness distribution varies across months, we standardize the measure to range from 0 and 1 within each month, where values close to zero represent directors with a low $k$ value and where the value 1 represents directors with the maximal value $k$. This variable expresses the continuous coreness of directors, with the value 0 indicating that directors are not in the largest component or not local brokers. Based on the largest component, local brokerage and coreness measures, we construct a discrete categorical variable that takes on the value 1 if directors are not in the largest component, 2 if they are in the largest component but not local brokers, 3 if they are in the largest component and they are local brokers, and 4 if they are in the innermost core (coreness = 1).

< Insert Figure 2 about here >

Figure 3 contrast the "inner circle" approach with our "network core" approach. The leftmost panels display results from a "conventional" analytical approach to detecting the "inner circle" in the context of Denmark's corporate network. Here we consider the network of director's based on the boards of the 500 largest companies. The "inner circle" has one large, connected component and 15 smaller components. The "network core" naturally consists of just one large component which by extension is much denser and therefore characterized by higher internal cohesion and greater overall reach centrality. The bottom panels highlight the distinctive nature of the two elite networks by emphasizing nodes in each network that also figure in the other. In the next section we go on to report on statistical analyses of the extent to which the core-periphery structure and company size is associated with the likelihood of joining a government committee.

---

[4] In the appendices (not submitted here but obtainable on request) we conduct sample sensitivity analyses of the $k$-core decompositions. Appendix Figure A1 shows that our analyses require a highly inclusive set of companies to identify the core-periphery structure, suggesting that the nested hierarchy of $k$-cores is structured by a wide substratum of director ties. Appendix Figure A2 shows that the decompositions are very robust to the exclusion of large companies. In Appendix Figure A3 we reproduce results altering the local brokerage threshold and Appendix Figure A4 shows the effect of altering the threshold on the efficacy of identifying political directors.

< Insert Figure 3 about here >

POLITICAL REPRESENTATION

*Political-institutional context*

The aim of our statistical analysis is to investigate the extent to which the network core capture corporate political representation in government committees, which is a key arena for business influence on policymaking and the legislative-regulatory process in the Danish parliament. Most prior work corporate networks and political representation is US-based and focuses on party donations, lobbying and positions on industry roundtables and/or policy commissions (Burris 2005; Heerwig and Murray 2019; Mizruchi and Hyman 2014; Murray 2017). We study Denmark, a coordinated market economy with a strong Social-Democratic variant of corporatist governance (Campbell and Pedersen 2007; Hall and Soskice 2001; Katzenstein 1985; Thelen 2014) centred on a mix of market- and state-based economic governance (Amin and Thomas 1996) which sets it apart from more state-centred variants such as Germany and France. On the one hand, the Danish economy is governed through decentralized associational decision-making with authority distributed across public and private institutions that interact via inter-institutional dialogue. Compared to other coordinated market economies such as Germany, the state has historically played a less dominant role in regulating the economy. Instead, direct negotiations between labor and capital have played a prominent role in policy and reform processes. For example, since the beginning of the 20th century labor and capital have pretty autonomously, and with only occasional interference by the state, reached binding agreements concerning the regulation of wages and other important political-economic issues (Jessop and Pedersen 1993; Nielsen and Pedersen 1991).

On the other hand, the Danish state followed the corporatist trajectory seen in a host of European countries during the post-war period. On matters of broader economic and industrial policy, the state develops policies via intermediation between umbrella interest organizations (Amin and Thomas 1996; Ibsen and Thelen 2017; Nielsen and Pedersen 1991). After the Second World War – the pinnacle of Denmark's corporatism – The Danish Trade Union Confederation (Landsorganisationen i Danmark) and Confederation of Danish Employers (Dansk Arbejdsgiverforening) exercised de facto decision-making power over most major policies and regulations in the Danish economy (Blom-Hansen 2001; Pedersen 2006), with shifting minority governments seeking input and support from the two interest organizations on major reforms related to wage moderation, employment policy, and fiscal policy (Lembruch and Schmitter 1982). At the same time a more decentralized web of permanent and ad hoc legislative committees and policy commissions with substantial everyday expert and interest group participation has since evolved to provide systematic inputs and legitimacy around the ongoing legislative process - both in the agenda-setting, policy formulation and implementation stage (Andersen, Dølvik, and Ibsen 2014; Anthonsen, Lindvall, and Schmidt-Hansen 2011; Due et al. 1994). While this system has been subject to minor ongoing reform, it has proved resilient and is still a key trait of democratic governance in Denmark.

Recent studies of interest group representation in the Danish political arena using survey data describe a system of "privileged pluralism", where multiple interests have access to the legislative process and opportunities to influence it, but where access and influence is unequally distributed and highly centralized around a few privileged interest groups, notably the prominent labour unions and business associations (Binderkrantz 2005; Binderkrantz and Christiansen 2015; Binderkrantz, Christiansen, and Pedersen 2015). The same interest groups have, together with big corporations and key actors from the central state administration, been identified as Denmark's "power elite" (AUTHOR(S)). Studies also find that private business interests in Denmark prefer lobbying legislation via formal corporatist channels and are less keen on informal lobbying and media-based agenda setting (Binderkrantz 2008). Previous research on the central circle of Denmark's largest corporations found that these directors frequently participate in policy-planning (AUTHOR(S)). However, no prior studies have systematically traced what representatives from the wider population of corporate elite individuals participate most intensely in democratic governance.

*Dependent variable*

We model the entire director networks' likelihood of acting on government committees as a function of their position in the core-periphery structure. Further, our analyses investigate if the political value of network coreness is derived mainly from the traits and resources of directors and the companies they are connected to, or if coreness in and of itself can predict if directors are committee members. Our dependent variable is membership of government committees which we identify using the Danish Elite Network Database (DEN). DEN was collected in the period 2013 to 2015 and again in the period 2016 to 2017. The data set consists of the names and addresses of the members of all 650 government committees active in the two periods. We conducted a name and address match between the list of committee members in DEN and the list of directors in the corporate governance registry (*Deltagerregisteret*). This match identified 597 committees with at least one director. For example, *the Competition Legislation Committee* (Udvalg om Konkurrencelovgivning) has 20 members that are also directors and provides advice on legislation to the *Ministry of Business Affairs*. About half of the members are public officials and legal experts, and about half represent business interests and labor. Six members are appointed directly by major business associations including the Confederation of Danish Industry (*Dansk Industri*) and Denmark's Business Association (*Dansk Erhverv*) and for these members executive business experience is required. For board directors that were active 2010 to 2012 we conducted a name and address match on the list of government committees' members active during 2013 to 2015, and for directors active 2013 to 2015 we repeated the match for members of government committees active 2016 to 2017. Of all active directors, we identify 3000 unique directors that are at some point a committee member.

We estimate the odds of government committee memberships using logistics regression. To comparability across different model specifications, we report the average marginal effects of the coefficients which can be interpreted as the predicted change in the probability of joining a committee given a one-unit change in the independent variable. Although our model is not identifying a causal relationship in the strict sense of the word and does not rule out endogeneity between corporate networks and political representation, two observations lead us to believe that the dominant causal pathway goes from corporate networks to committee memberships. First, Useem's theory of the inner circle (1986) laid out a historical process where the formation of a cohesive elite network of business leaders from the large corporations preceded their ascendance as a collective political agent in American politics. Second, and more importantly, the institutional-political context that this paper is investigating reinforce this interpretation. As described above, members of government committee that are not public officials or fixed members of business and trade associations are formally selected due to their leadership and industry experience and status (their position in the corporate network likely expresses this). This means that we would expect this experience to precede committee selection. Nevertheless, we cannot rule out cases of directors that are well-connected due to their experience, but are not in the network core, who then are selected to act on a committee which affords them with new connections that lead to additional board recruitment, which then in turn moves them into the core. Even if such reverse causality may be at play to some extent, the model is still useful for assessing the pathways between corporate elite networks and democratic governance, and for investigating the extent to which director and firm characteristics moderate these pathways. It is these characteristics that we now turn to.

*Independent variables*

Core-periphery network structure: We construct a categorical measure that classifies the population of directors into four discrete categories ranging from least to most elite. The first category consists of directors that in each month are not integrated in the largest component. The second category consists of redundant directors from the largest component that did not qualify as local brokers. The third category consist of directors that qualified as local brokers but did not qualify as maximally core, thus situated somewhere in the semi-periphery of the network of brokers. The fourth category consists of directors that qualified as belonging to the innermost k-core.

Company size, director roles and the "inner circle": All existing research on corporate elite networks emphasize the role of company size, but because this work focuses solely on large corporations there is very little work that explores variability in company size, network structure and political representation (Huijzer and Heemskerk 2021). To capture these relationships, we calculate an annual composite rank based on three company size indicators: number of employees, total revenue and total assets. To do so we draw on accounting data from Statistics Denmark's company registry (*FIRM*). We apply Principal Component Analysis (PCA) to the population of companies in our data and order them by rank from top to bottom according to the first dimension identified in the PCA. The first dimension of the PCA captures 68% of the total between-company variation in the three variables. Since directors can have multiple board position, each director is assigned the rank of their highest ranked company in the board portfolio. Figure 4 plots the cumulative share of total employment, revenue and assets along the rank distribution described above, and illustrates the concentration of economic resources around the largest companies Denmark. We classify director company rank into four categories: "Top 1-50"; "Top 51-500", "Top 501-5000" and "Top 5000-". We performed an equivalent ranking of *corporations* based on the three variables aggregated across *companies*, where a nominally small company gets the rank of its corporation.[5] We used the corporation registry (*Koncernregistret*) which links all subsidiaries in FIRM to its ultimate parent. We also ran statistical analyses based on this corporation rank measure, and results are extremely similar to the main results.

Next, we coded indicator variables for director roles as Chair or Executive (whether a director in at least one "Executive" role and one "Chair" role) and also classified director roles by company size. To approximate existing operationalizations of the "inner circle" (Useem 1986) we coded dummy variables for top linkers, that is, directors with multiple positions in top 50 and top 500 companies (see also Figure 3 above).

< Insert Figure 4 about here >

<u>Tripartite interest group actors:</u> As mentioned above, Denmark's social-corporatist mode of governance is baked into the government committee system. Above and beyond elected politicians, committees consist of standing members from business associations, trade union and/or public servants. To absorb the increased odds of being selected to committees for directors who also hold high level positions in any of these three organizations, we identify the primary job of all directors from the employment registry (*RAS*). We include binary indicators for directors that currently are, or previously have been elected politicians, leaders of business associations and/or leaders of labor unions.[6] We also control for business association committee membership recorded during the 2012 to 2017 period drawing on DEN, since these members are also likely to have standing roles in government committees due to their interest group affiliation. Of the 414 business association committees consisting of 6122 unique individuals figuring in DEN, 397 consists of at least one match from our director sample. For example, *The Executive Committee at Confederation of Danish Industries* (Forretningsudvalget, Dansk Industri) consists of 33 corporate representatives. Overall, we find that of the 130,000 directors in our sample, 1300 figure on business association committees after performing our name and address match procedure.

---

[5] In this paper we distinguish between *company* (the legal entity registered for regulatory and accounting purposes, in Danish *virksomhed*) and *corporation* which consists of the corporate group (in Danish *koncern*) including parent company and all subsidiaries.

*Control variables*

Company controls: For a directors' highest ranked board, we distinguish subsidiaries and publicly traded companies with indicator variables, company age with a categorical variable ("1-11 years", "12-25 years", "26-50 years", "50 and above years"), and a categorical variable for industry (1-digit NACE).[7]

Director controls: We control for directors' demographic characteristics by including an indicator for female directors, a categorical migrant status variable ("Native", "Immigrant" and "Descendant of immigrant"), and a categorical variable for director age ("18-30 years". "31-44 years", "45-59 years", "60-74 years". "75 and above years"). We also control for the broad socio-economic status of directors through indicators of college and master level degrees, binary indicators for whether directors are in the top 0.1 percentile of the national income and wealth distribution, and finally a categorical class origin measure that indicates the highest occupation of a directors' parents ("Employer", "Manager", "Professional", "Other").

---

[6] In the main models we only restrict this measure to ISCO-1 managers in business associations and unions. We ran alternative models where we include all employees in business associations or unions and results are extremely similar. Elected politicians we identify from the election database (VALG) which record all elected politicians dating back to 1992.

[7] Again, we ran additional analyses where all subsidiaries were coded according to the parent company's characteristics, and estimates were extremely similar.

*Is the network core politically distinct from the inner circle?*

Table 1a and 1b contrasts directors (and linkers) among respectively top 50 and top 500 companies and the network core. The network core is populated by (a monthly average of) 278 directors. A major part of the network core is affiliated with a large corporation. 80 percent of all its directors sit on the board of a company that has a top 500 company as the ultimate owner, and 38 percent sit on the boards of a company that has a top 50 corporation as the ultimate owner. A significant proportion of the network core however consists of directors in small companies. 20 percent consists of directors entirely unaffiliated with large corporations. 35 percent do not hold a position in a top 500 company, and 80 percent do not hold a position in a top 50 company. A significant proportion of the directors in the network core are therefore either directors in smaller independent companies or in smaller subsidiaries.

< Insert Table 1 about here >

Are connections among top directors a good proxy for network core membership? Table 1A and 1B shows that most directors in top companies sit on multiple top boards. Most of these so-called linkers however connect smaller subsidiaries within or between large corporations. Just 11 percent of all directors on the board of a top 50 company (256) sit on the board of multiple top 50 companies (29); and just 15 percent of all directors on the board of a top 500 company (1899) sit on multiple top 500 boards (294). These linkers make up a minority of the network. While about half of the linkers between top 50 companies (29) are also in the network core (13), they make up just five percent of the overall network core (278). About 30 percent of the linkers between top 500 companies (294) are also in the network core (87), and they make up 30 percent of the network core. If we define top linkers more broadly as consisting of directors that sit on any company (subsidiary or parent) within a top corporation (corporate group), we find that the network core consists of respectively 78 percent top 500 linkers and 36 percent top 50 linkers. However, 90 percent of the overall population of broad company-based top 50 linkers, and 95 percent of the broad firm-based top 500 linkers, are not in the network core. In sum, while linkers of top companies and corporations are more likely to enter the network core than non-linkers, the network core represents a distinct subset of directors that cannot just be inferred from the connections among top directors.

Tables 1A and 1B also show the distinct political characteristics of top directors and the network core. The network core stands out as distinctly political. 19 percent of the directors in the network core (278) sit on government committees (52), 18 percent are previous leaders of business associations (50), and nine percent are members of business association committees (25). We also observe six percent previously elected politicians (17) and previous leaders of labor unions (16) which indicate that their political capital is an asset for companies and likely provide the business community with valuable political connections.[8] The overall proportion of political directors among top companies is lower even when it comes to the linkers (typically considered "the inner circle"). In sum, the network core stands out as consisting of directors with distinctly political attributes, especially those in the network core that are also executives in or linkers between top 50 companies. When it comes to members of government and business association committees' executives of and linkers between top 50 companies is the group that comes closest to the network core in terms of relative representation, yet these directors represent a markedly smaller group and do not to the same extent embed powerful actors from the social-corporatist system that characterize Denmark's democratic governance.

---

[8] Current politicians and leaders of interest organization figure less frequently, which is likely due to norms about impartiality linked to such political roles. Nevertheless, elected politicians occasionally serve on the boards of firms where public authorities have some direct ownership or indirect stakes.

In Figure 5 we visualize the concentration of political directors across the coreness distribution in the overall director network. In panel 4.A we see that the coreness structure is particularly strongly associated with an increase in the leaders of business associations and labor unions, which is tripled when moving from the set of directors with no coreness to directors with maximum coreness (the network core). In panel 4.B and 4.C we see that the proportion of directors in government committees and business committees also increase substantially as their coreness goes up. The panels all show that the concentration gradient becomes steeper the closer directors are to the innermost core of the network.

< Insert Figure 5 about here >

Model results: To identify the relative strength of coreness and company rank in predicting government committee membership we now present results from our logistic regressions. To reiterate, Table 2 reports marginal effects from the model results. This eases comparison of estimates across models and can be interpreted as the change in the predicted probabilities of joining a government committee associated with a one-unit change in the independent variable. Model 1 is a univariate model that contains only the elite network variable. This model estimates change in the predicted probability of joining a government committee as a function of moving from outside the largest component into respectively the network core, to being a local broker and to a position in the largest component. The mean probability of being member of a government committee across the analytical director sample is 0.007, or 0.7 percent, across the periods. If a director moves from outside the largest component into the network core, we expect a 24-percentage point increase in probability.

Local brokers are also significantly advantaged by an increase in probability of 6 percentage points. Directors in the largest component are significantly less likely to join a government committee than local brokers or directors in the network core, but still about twice as likely as directors outside the largest component with an increase of about 0.7 percentage points. These estimates are reduced somewhat but all remain large and significant when controlling for the number of boards positions a director has in model 2.

< Insert Table 2 about here >

Model 3 estimates the probability increases of top linkers. This model shows that directors who sit on multiple top 50 boards are as advantaged as directors in the network core. However, this group of directors is almost 10 times smaller than the network core (29 vs. 278, see Table 1). Model 4 shows that the top linker advantage is almost fully accounted for by adjusting for the position core-periphery structure. Model 5 focuses on company rank. This model estimates a strong increase of about 15 percentage points for top 50 directors over directors of small boards (> top 5001, which is the reference category) and we also find a strong though diminishing increase for top 500 and top 5000 directors. Model 6 shows that increases remain positive and significant even after accounting for the rank of corporate groups. Model 7 focuses on director roles net of rank and top linker status and shows that executives and chairs of top boards are somewhat advantaged over ordinary board members (both with an increase of 0.8 percentage points). This model also shows that the top linker measures are almost fully accounted for by rank and role: The model shows that being a top 50 linker contributes only with a 0.2 percentage point increase, or about 1/3 of the baseline probability. These estimates indicate that the economic resources of companies help propel directors into government committees, but that connections across top ranked board adds little advantage net of company rank.

Model 8 estimates the independent probability increase of elite network and firm rank together. The elite network estimates remain very strong and positive with an increase of about 10 percentage points. Contrasting with model 1, we see that about 60 percent of the advantage stemming from elite networks is accounted for by rank. The rank measure also remains strong but significantly less strong than the elite network measure, with an increase of about 2 percentage points for top 50 directors. Contrasting model with model 5, we see that about 85 percent of the advantage stemming from rank is accounted for by elite networks, a higher proportion than vice versa.

Model 9 includes all elite network, top linker, rank and role measures simultaneously, showing that the strongest independent increase stems from the network core and the local brokers. These estimates are entirely robust to the inclusion of fixed effects for a director's number of boards (see model 10), indicating that network coreness captures elite characteristics above and beyond degree centrality. Models 11 and 12 introduce company and director controls separately and model 13 includes them simultaneously. In these models the network core estimates remain strong and significant. Contrasting models 10 and 11 with model 9, we see that company controls account for more of the independent effect of network core in model 9 than director controls do. Both variable classes are nevertheless strong mediators of the network core advantage, suggesting that both firm types and socio-demographic as well as broader class characteristics are important explanations of political advantage in the corporate elite.

Policymaking in Denmark is heavily influenced by social-corporatist institutions where capital and labor interests are baked into the system of governance. As we saw in the descriptive analysis above (see Table 1), this structure is reflected in the composition of the network core. Table 3 presents estimates from models accounting for leading roles in interest groups. Models 14 through 17 introduce different roles separately and model 18 includes all roles simultaneously. Contrasting these models with model 10, we see that leading roles in interest organization reduce the network core estimate additionally (by 20 to 25 percent). Not surprisingly the variables adjusting for leadership roles in business associations is the strongest mediator. This suggest that one pathway into government committees for the network core is through representation of business associations. We also see from these models that directors from interest organizations are strongly advantaged net of elite network and company rank. In the final model 19 we include all variables and controls. This model explains about 1/3 of the overall variation (as measured by the pseudo R2). In this model the effect of network core is substantially reduced to a 2.5 percentage point increase, still more than three times higher than the baseline probability, but ten times less than in the initial univariate model (model 1).

< Insert Table 3 about here >

Coreness-company rank interactions: Network coreness and rank both strongly predict if directors join government committees. Does network coreness compensate the initial political disadvantage of directors that are not in top ranked firms, or do directors of large corporations located in the network core experience a double advantage? Figure 5 presents the marginal effects from models interacting the continuous coreness measure with company rank. Again, we calculate change in predicted probability but now differentially for directors of varying company rank and at coreness values ranging from zero to one (0, 0.25, 0.50, 0.75 and 1). The top and mid panels of the figure show that directors of all ranks benefit substantially from coreness, but directors of smaller companies benefit significantly more from coreness than directors of large companies – even when considering the aggregated rank of the entire corporation. In fact, the total advantage for directors of very small companies exceeds that of directors in larger companies (or corporations) once they are part of the network core (coreness=1). This suggests that network coreness fully compensates for the lesser economic resources among directors of smaller companies, and that in fact directors of larger companies are somewhat penalized by their high rank. The lower panel presents results from a model including a three-way interaction between coreness, company rank and corporate group rank. Based on this model we calculated the advantage for small companies (rank >5000) across the coreness and corporate group rank distribution. The plot shows that directors of small subsidiaries within large corporate groups enjoy a distinct advantage suggesting that these directors play a surprisingly strategic role as political representatives for large corporations.

< Insert Figure 6 about here >

CONCLUDING DISCUSSION

Previous research on the political representation of corporate elites has focused on the network structure of directors connecting a country's largest corporations (Burris 2005; Comet 2019; Heemskerk and Fennema 2009; Mills 1956; Mizruchi 2013; Useem 1986). We proposed a method that simultaneously capture the network qualities of brokerage, centrality and cohesion: a two-step network decomposition that locates the coreness of local brokers. Analyzing the population network of all Danish company directors, we identify a core-periphery that emerge from connections among a wide distribution of directors in companies of varying size. We also show that directors with positions in the network core are more advantaged than directors of large companies. Our analyses further show company rank is associated with significant advantages, but that the network core can fully compensate for the disadvantage of not having high-ranked positions. Finally, the approach also locates leaders of key interest organizations, including industry associations and labor unions, that make up the social parties of social-corporatist Denmark. An approach focusing exclusively on the network among directors of high-ranking boards, in contrast, captures a significantly lower fraction of directors engaged in business political activity and with representation in government committees.

Our results have implications for studies of corporate elite networks. Sampling strategies matter greatly in how well elite networks are measured and with increased access to large-scale network data from administrative data measurement will likely improve. Our study shows that network connections at the tail end of the company rank distribution contribute to elite cohesion and integration of corporate elites both higher-ranking and lower-ranking companies. The analyses also point to the potential of using large-scale network data to identify powerful brokers within the complex networks of large corporate groups, which are not necessarily top executives, as well as smaller elite companies that are nominally not as well-resourced. Corporate restructuring in market economies around the world warrant researchers of corporate elites to develop new methods of capturing key actors across different political-economic contexts. Comparative research should seek to establish if corporate elites in other countries too are organized in a network core of brokers and if not what scope conditions may apply.

Several scope conditions of Denmark's social-corporatist economy the influence of broader business interests. Albeit more and more economic resources concentrate around a few large Danish corporations (such as Novo Nordisk, Mærsk and Danske Bank), Denmark still have relatively few large companies and top-500 ranked companies employ as little as 200 employees. The economy therefore rests on a significant component of small-to-medium sized (particularly export oriented) companies that play an important role in Denmark's economic performance (Kristensen and Sabel 1997). Denmark's political institutions are in some measure designed to give voice to a wide network of agricultural smallholders. This economic structure likely gives way to a principle of elite organizing that hinge more strongly on local connections among smaller units, but which also selects local representatives into field-level political processes (Boberg-Fazlic et al. 2023). Denmark's corporatist system of governance provides many entry points for corporate interests to have voice in policymaking and the legislative process. This possibly opens space for a wider and more heterogenous set of actor interests in the exercise of corporate political influence. Our paper shows that above and beyond interest organizations, the network core is a key point of selectivity in how these interests become connected with policymaking. Whether actors from the wider business community are equally influential in contexts of larger corporate concentration and in institutional contexts with less entry points into politics should be investigated. Few comparative studies exist on variability in corporate elite organizing (for an exception see Cárdenas (2012)) and future work should consider the role of the wider business community.

While our findings show that the economic resources of companies together with the structure of wider elite networks together determine access to political influence, the independent effect of elite networks may to some extent be explained by the existence of this social-corporatist consensus model. Perhaps directors with prominent positions in the core-periphery structure of the wider business community have a higher degree of public legitimacy, than the faces of large corporations perceived as representing vested interests? Nevertheless, large companies send less prominent actors from their internal network into positions of political influence. How restructuring of the corporate landscape affects elite networks is an important topic requiring detailed analyses of longitudinal data. Going forward studies of corporate elites would benefit from more micro-level studies on the internal network structure of large corporations (corporate groups) and whether strategic positions within these networks afford directors with social capital to exercise influence within society and politics more broadly. Large corporations may wish to delegate different political roles across the complex network organizations depending on business strategic considerations.

**Figure 1.** Fraction of the largest component iteratively identified as local brokers

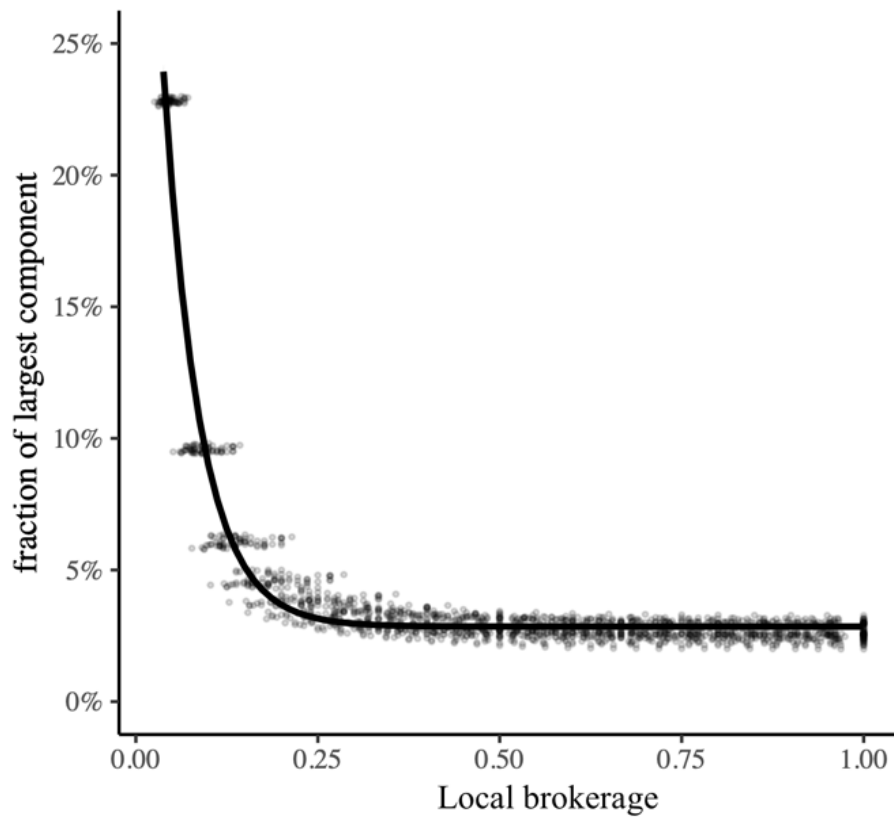

Note: The figure plots the fraction of directors in the largest component that remain at each step of the local brokerage pruning.

**Figure 2.** Results from k-core decomposition.

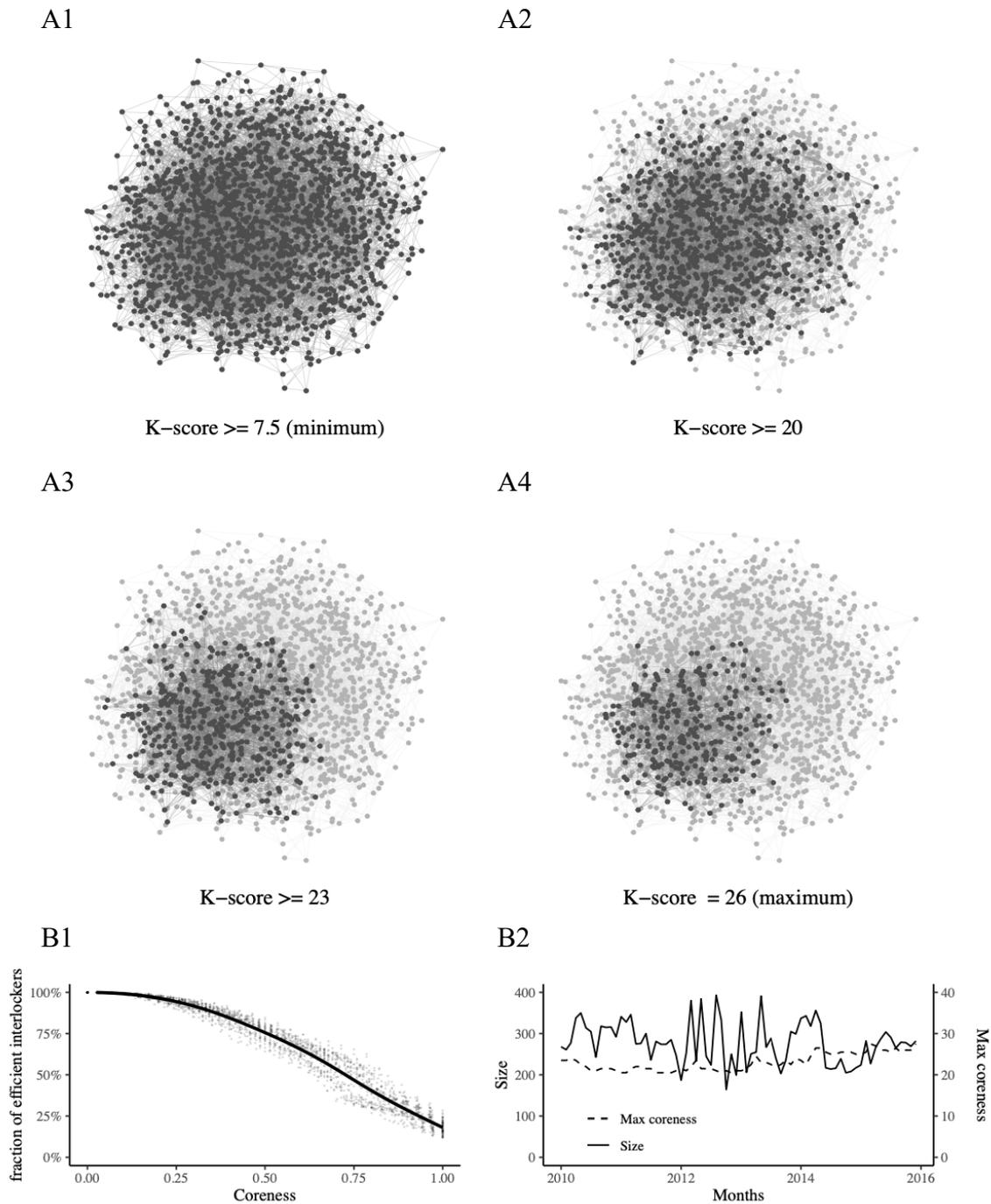

Note: Panels A1-4 illustrates the step-wise k-core decomposition. Panel A1 plots the full network of local brokers in October 2015 using the default setting in the Fruchterman-Rheingold algorithm of the Igraph R-package. Panels A2 to A4 highlight the directors remaining at select k-steps in the decomposition (from low to max), with panel A4 stressing the maximal k-core. Figure B1 reports the average fraction of efficient interlockers remaining at each standardized coreness value across the months. Figure B2 plots the number of nodes *i* with maximal *k* by each month.

**Figure 3.** Director network illustrations. The "inner circle" vs. the k-core approach

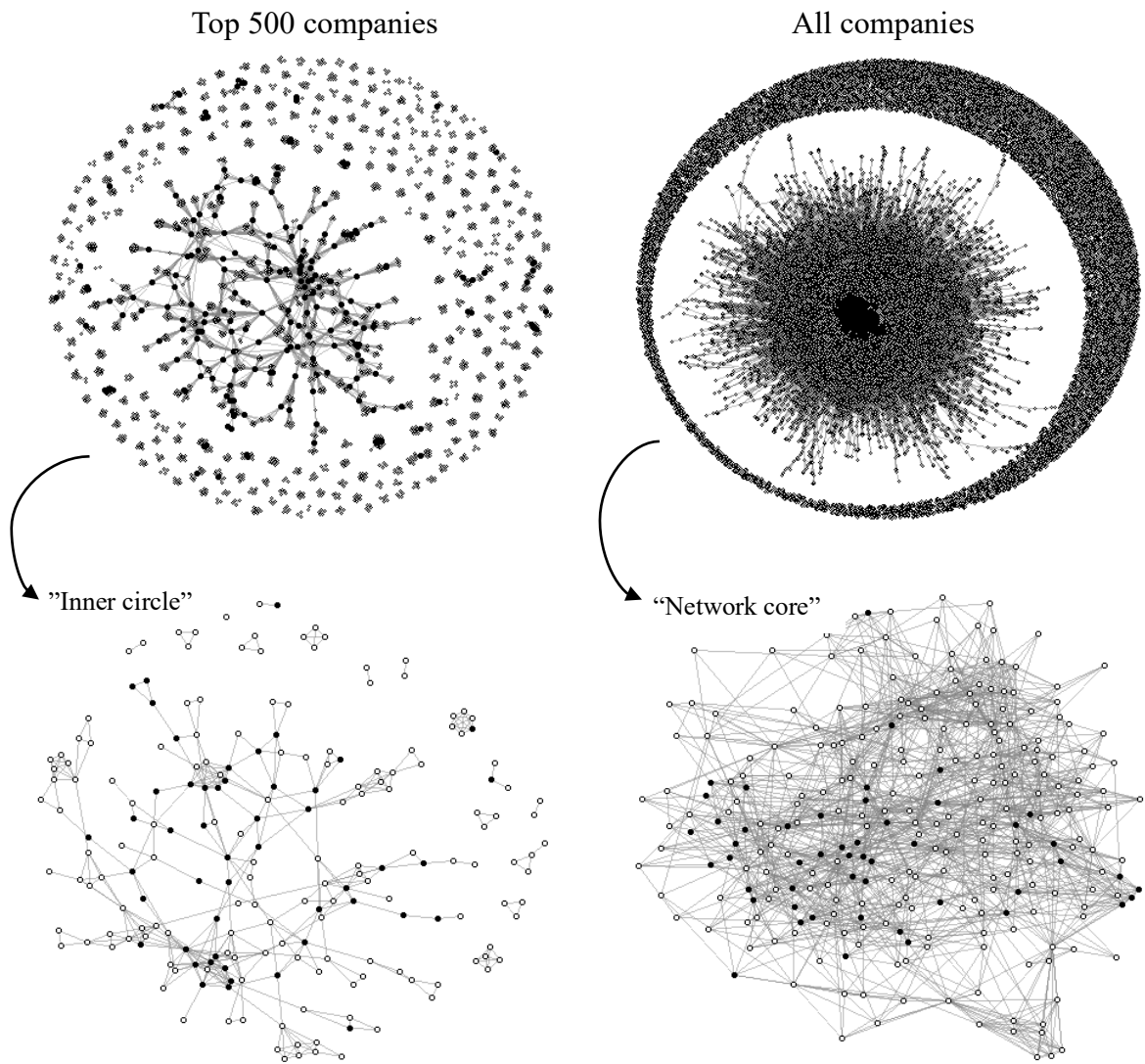

Note: The figure contrasts the "inner circle" approach to elite detection with the k-core approach. In the leftmost panels we display an analytical process that begins from the full director network of the top 500 companies and ends with the "inner circle" which is the subset of director with multiple board positions (emphasized in black in the full network of top 500 directors). The rightmost panels begin with the full director network of all companies in the population sample and end with the network core, which is the subset of directors with the maximal k-core score (emphasized in black in the full network of all directors). In the bottom panels we highlight the contrast between the "inner circle" and the "network core" by emphasizing nodes in black that are in the other subset. For the "inner circle" illustration we show directors that are also in the core in black. For the "network core" illustration we show directors that are also in the inner circle in black. All networks displayed are for the month of October 2015.

**Figure 4.** Corporate concentration and company rank in Denmark's economy.

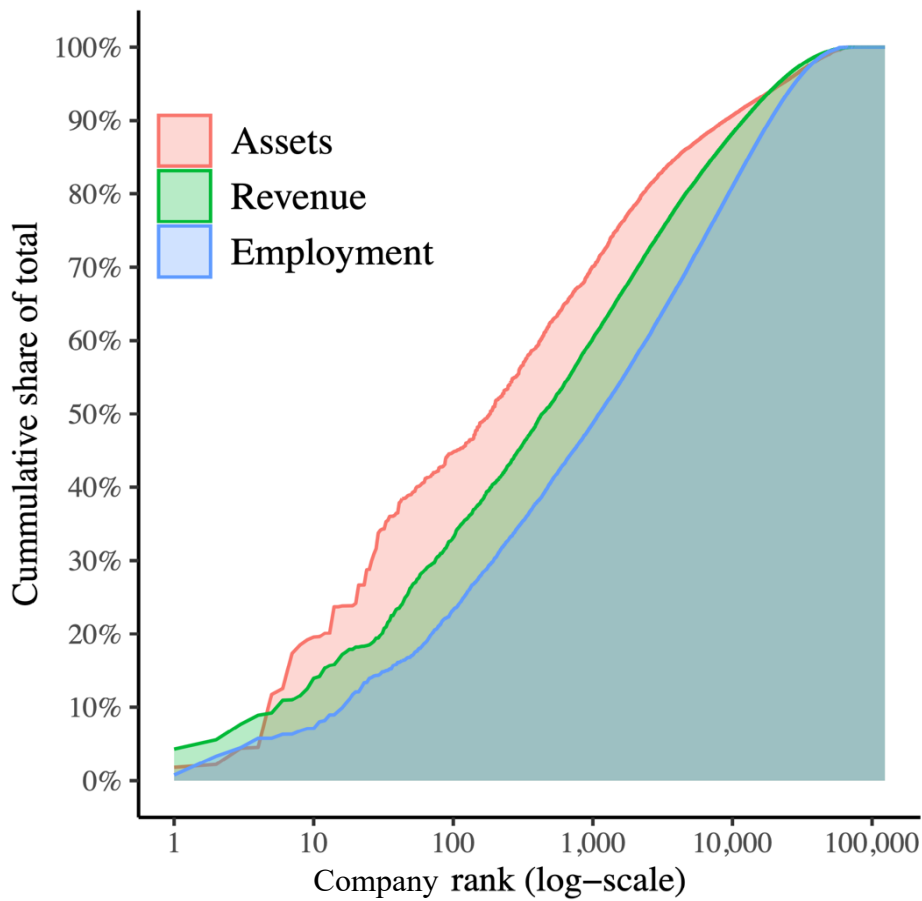

Note: The figure reports the cumulative share of assets, revenue and employment for all companies from top to bottom ranked companies (for details on the rank measure see description of the PCA-ranking method above).

**Figure 5.** The concentration of political directors in the core-periphery structure.

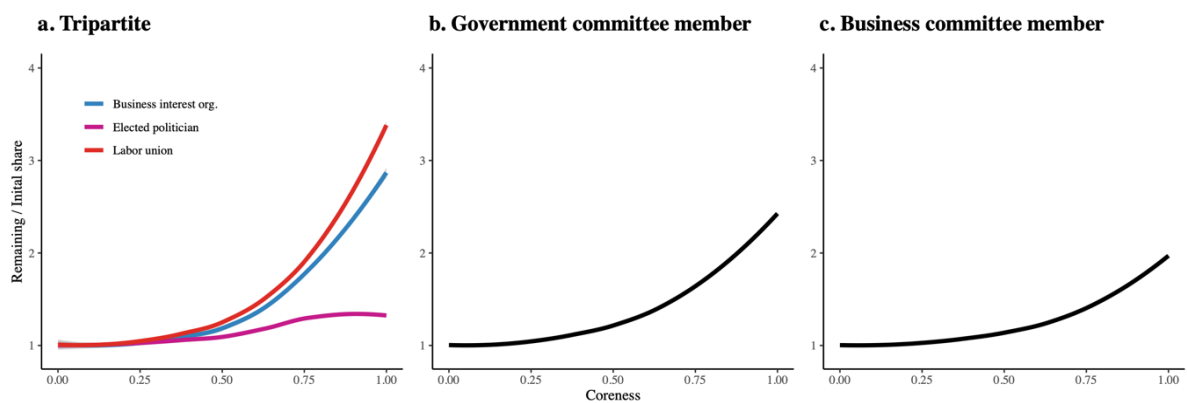

Note: The figure plots the relative increase in the share of directors at each value of standardized coreness. The baseline share is calculated on the full sample of efficient directors. The line represents the average over all 72 months.

**Figure 6.** Predicted change in probability of government committee over rank.

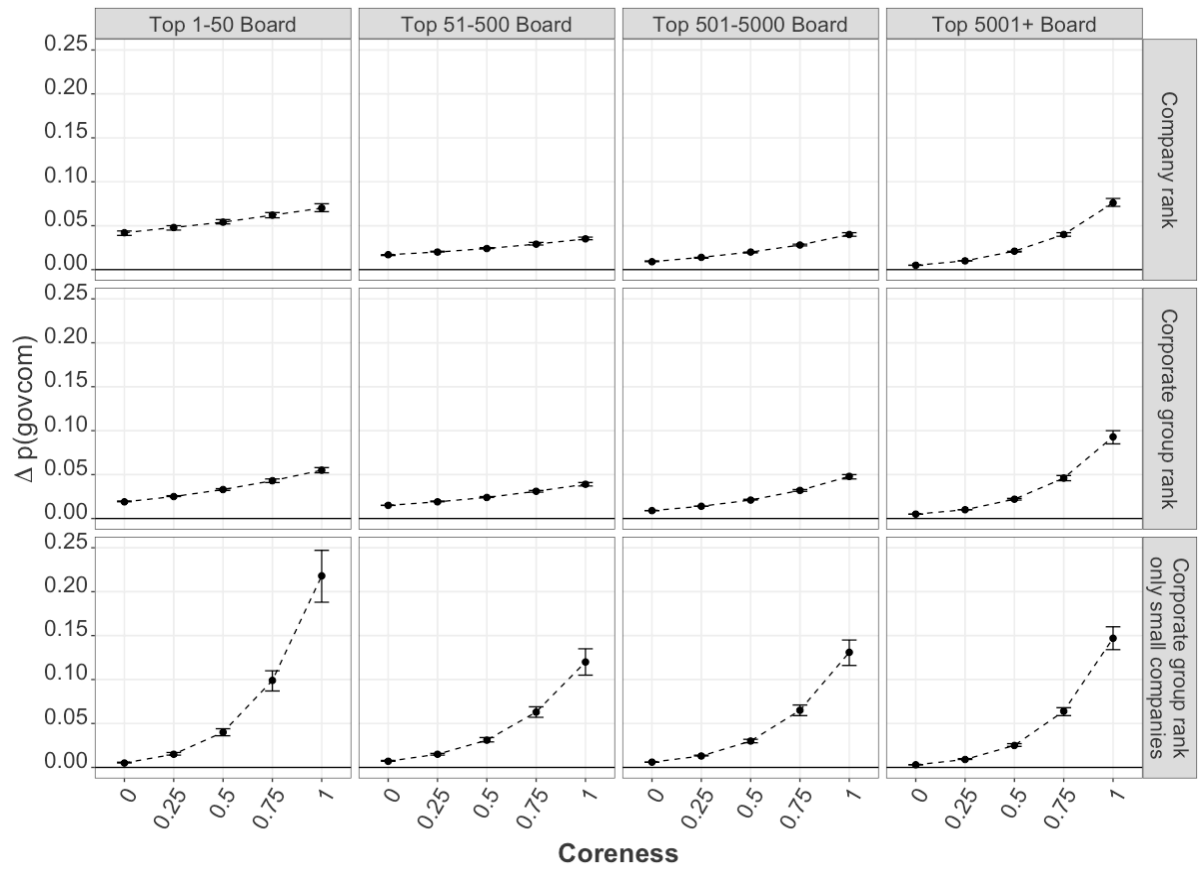

Note: Prediction from full model with interaction terms. Error bars represent 95% confidence intervals.

## Table 1a: Top 50 and the network core

| | | Director type: | | | | Linker type: | | | |
|---|---|---|---|---|---|---|---|---|---|
| | All | Within top 50 corp. group | Within top 50 company | Executive within top 50 company | Chair within top 50 company | Between subs. of top 50 corp. group | Between top 50 companies | Between subs. across top 50 corp. groups | Between subs. within top 50 corp. group |
| **Total directors** | 149691 | 1096 | 256 | 111 | 50 | 1064 | 29 | 484 | 1062 |
| Share of directors: | | | | | | | | | |
| Business committee | 0.5% | 4.8% | 9.0% | 14% | 20% | 4.5% | 10% | 6.2% | 4.5% |
| Government committee | 0.4% | 4.7% | 12.5% | 17% | 6.0% | 4.3% | 21% | 5.8% | 4.3% |
| Current leader of business association | 0.1% | 0.3% | 0.4% | 0.0% | 0.0% | 0.02% | 0.0% | 0.0% | 0.02 |
| Previous leader of business association | 1.5% | 5.7% | 8.6% | 8.1% | 8.0% | 5.5% | 6.9% | 7.0% | 5.5% |
| Current leader of labor union | 0.1% | 0.0% | 0.0% | 0.0% | 0.0% | 0.0% | 0.0% | 0.0% | 0.0% |
| Previous leader of labor union | 0.5% | 0.6% | 0.8% | 0.0% | 2.0% | 0.6% | 3.4% | 0.6% | 0.6% |
| Current elected politician | 0.6% | 0.6% | 0.4% | 0.0% | 0.0% | 0.7% | 0.0% | 0.6% | 0.6% |
| Previous elected politician | 1.3% | 2.0% | 2.0% | 0.0% | 0.0% | 1.9% | 0.0% | 2.1% | 1.9% |
| **Network core** | 278 | 106 | 58 | 20 | 21 | 100 | 13 | 71 | 100 |
| Share of network core: | | | | | | | | | |
| Business committee | 9.0% | 8.5% | 8.6% | 20% | 29% | 9.0% | 15% | 9.9% | 9.0% |
| Government committee | 19% | 19% | 21% | 25% | 5.0% | 18% | 23% | 17% | 18% |
| Current leader of business association | 1.7% | 0.1% | 1.7% | 0.0% | 0.0% | 0.0% | 0.0% | 0.0% | 0.0% |
| Previous leader of business association | 18% | 16% | 17% | 15% | 10% | 15% | 7.7% | 14% | 15% |
| Current leader of labor union | 1.7% | 0.0% | 0.0% | 0.0% | 0.0% | 0.0% | 0.0% | 0.0% | 0.0% |
| Previous leader of labor union | 5.7% | 1.9% | 1.7% | 0.0% | 5.0% | 2.0% | 7.7% | 1.4% | 2.0% |
| Current elected politician | 4.3% | 1.0% | 1.7% | 0.0% | 0.0% | 1.0% | 0.0% | 1.4% | 1.0% |
| Previous elected politician | 6.1% | 2.8% | 3.4% | 0.0% | 0.0% | 3.0% | 0.0% | 1.4% | 3.0% |

## Table 1b: Top 500 and the network core

| | | Director type: | | | | Linker type: | | | |
|---|---|---|---|---|---|---|---|---|---|
| | All | Within top 500 corp. group | Within top 500 company | Executive within top 500 company | Chair within top 500 company | Between subs. of top 500 corp. group | Between top 500 companies | Between subs. across top 500 corp. groups | Between subs. within top 500 corp. group |
| **Total directors** | 149691 | 4549 | 1899 | 756 | 392 | 4139 | 294 | 2074 | 4107 |
| Share of directors: | | | | | | | | | |
| Business committee | 0.5% | 4.4% | 6.7% | 9.8% | 7.1% | 4.2% | 9.5% | 5.9% | 4.2% |
| Government committee | 0.4% | 3.6% | 6.2% | 6.6% | 9.1% | 3.4% | 11% | 4.3% | 3.4% |
| Current leader of business association | 0.1% | 0.3% | 0.5% | 0.0% | 0.3% | 0.3% | 0.3% | 0.3% | 0.3% |
| Previous leader of business association | 1.5% | 6.0% | 8.7% | 6.8% | 13% | 5.6% | 12.6% | 6.3% | 5.5% |
| Current leader of labor union | 0.1% | 0.2% | 0.04% | 0.0% | 0.5% | 0.03% | 0.03% | 0.01% | 0.03% |
| Previous leader of labor union | 0.5% | 1.3% | 1.5% | 0.5% | 1.8% | 1.3% | 1.7% | 1.1% | 1.3% |
| Current elected politician | 0.6% | 2.3% | 4.5% | 0.0% | 7.1% | 1.3% | 5.1% | 1.3% | 1.0% |
| Previous elected politician | 1.3% | 3.9% | 6.8% | 0.0% | 12% | 2.7% | 7.5% | 2.6% | 2.3% |
| **Network core** | 278 | 224 | 179 | 62 | 79 | 217 | 87 | 180 | 216 |
| Share of network core: | | | | | | | | | |
| Business committee | 9.0% | 9.4% | 10% | 15% | 8.8% | 9.7% | 10% | 9.4%% | 9.7% |
| Government committee | 19% | 18% | 18% | 18% | 20% | 18% | 20% | 17% | 18% |
| Current leader of business association | 1.8% | 1.3% | 1.1% | 0.0% | 0.0% | 1.4% | 1.1% | 1.1% | 1.4% |
| Previous leader of business association | 18% | 16% | 17% | 13% | 16% | 16% | 16% | 16% | 16% |
| Current leader of labor union | 1.7% | 1.3% | 1.7% | 0.0% | 1.3% | 1.4% | 1.1% | 0.6% | 1.4% |
| Previous leader of labor union | 5.7% | 4.0% | 4.5% | 1.6% | 3.8% | 4.1% | 2.3% | 2.8% | 4.2% |
| Current elected politician | 4.3% | 2.2% | 2.2% | 0.0% | 3.8% | 1.8% | 1.1% | 1.7% | 1.9% |
| Previous elected politician | 6.1% | 4.5% | 4.5% | 0.0% | 6.3% | 3.2% | 2.3% | 2.2% | 3.2% |

**Table 2. Predicted change in probability of government committee**

| | (1) | (2) | (3) | (4) | (5) | (6) | (7) | (8) | (9) | (10) | (11) | (12) | (13) |
|---|---|---|---|---|---|---|---|---|---|---|---|---|---|
| Elite Network (ref=Not in largest component): | | | | | | | | | | | | | |
| *Network core* | .239*** | .199*** | | .205*** | | | | .097*** | .101*** | .104*** | .056*** | .068*** | .045*** |
| | (.233 - .245) | (.192 - .205) | | (.198 - .211) | | | | (.093 - .101) | (.096 - .106) | (.099 - .109) | (.053 - .058) | (.064 - .071) | (.042 - .047) |
| *Local brokerage* | .063*** | .051*** | | .060*** | | | | .036*** | .036*** | .037*** | .028*** | .028*** | .024*** |
| | (.061 - .064) | (.049 - .053) | | (.059 - .062) | | | | (.034 - .037) | (.034 - .037) | (.036 - .039) | (.027 - .030) | (.027 - .029) | (.023 - .025) |
| *Largest component* | .007*** | .006*** | | .007*** | | | | .006*** | .006*** | .006*** | .006*** | .005*** | .005*** |
| | (.007 - .007) | (.006 - .006) | | (.007 - .007) | | | | (.006 - .006) | (.006 - .006) | (.006 - .006) | (.005 - .006) | (.005 - .005) | (.005 - .005) |
| Top 50 linker (=1) | | | .238*** | .010*** | | | .002*** | | -.002*** | -.002*** | | | |
| | | | (.219 - .256) | (.008 - .013) | | | (.001 - .003) | | (-.003 - -.002) | (-.003 - -.002) | | | |
| Top 500 linker (=1) | | | .114*** | .003*** | | | .001*** | | -.002*** | -.002*** | | | |
| | | | (.109 - .119) | (.003 - .004) | | | (.001 - .002) | | (-.003 - -.002) | (-.002 - -.002) | | | |
| Top rank board (ref=Top 5001-): | | | | | | | | | | | | | |
| *Top 50* | | | | | .145*** | .027*** | .060*** | .023*** | .012*** | .010*** | .015*** | .007*** | .013*** |
| | | | | | (.140 - .150) | (.025 - .028) | (.056 - .065) | (.021 - .024) | (.011 - .014) | (.009 - .012) | (.013 - .016) | (.006 - .008) | (.011 - .014) |
| *Top 51-500* | | | | | .048*** | .006*** | .031*** | .010*** | .006*** | .006*** | .009*** | .005*** | .007*** |
| | | | | | (.047 - .049) | (.006 - .007) | (.030 - .033) | (.009 - .010) | (.006 - .007) | (.005 - .006) | (.008 - .009) | (.004 - .005) | (.007 - .008) |
| *Top 501-5000* | | | | | .010*** | .001*** | .010*** | .003*** | .003*** | .003*** | .004*** | .002*** | .003*** |
| | | | | | (.009 - .010) | (.001 - .001) | (.009 - .010) | (.003 - .003) | (.003 - .003) | (.002 - .003) | (.004 - .004) | (.002 - .002) | (.003 - .004) |
| Top 50 executive (=1) | | | | | | | .008*** | | .012*** | .012*** | .010*** | .009*** | .008*** |
| | | | | | | | (.007 - .009) | | (.010 - .013) | (.010 - .014) | (.009 - .012) | (.007 - .010) | (.007 - .010) |
| Top 50 chair (=1) | | | | | | | .008*** | | .003*** | .004*** | .003*** | .002*** | .003*** |
| | | | | | | | (.007 - .010) | | (.002 - .004) | (.003 - .005) | (.002 - .004) | (.001 - .003) | (.002 - .004) |
| Top 500 executive (=1) | | | | | | | .002*** | | .003*** | .002*** | .002*** | .001*** | .002*** |
| | | | | | | | (.001 - .002) | | (.003 - .003) | (.002 - .003) | (.002 - .003) | (.001 - .002) | (.001 - .002) |
| Top 500 chair (=1) | | | | | | | .010*** | | .006*** | .006*** | .006*** | .005*** | .005*** |
| | | | | | | | (.009 - .011) | | (.005 - .006) | (.006 - .007) | (.005 - .006) | (.004 - .005) | (.004 - .006) |
| Director-month observations | 9,466,836 | 9,466,836 | 9,466,836 | 9,466,836 | 9,466,836 | 9,466,836 | 9,466,836 | 9,466,836 | 9,466,836 | 9,466,836 | 9,466,836 | 9,466,836 | 9,466,836 |
| Pseudo R2 | .123 | .132 | .019 | .124 | .076 | .091 | .080 | .141 | .143 | .151 | .186 | .185 | .211 |
| Number of boards FE | No | Yes | No | No | No | No | No | No | No | Yes | Yes | Yes | Yes |
| Corporate group rank | No | No | No | No | No | Yes | No | No | No | No | No | No | No |
| Company controls | No | No | No | No | No | No | No | No | No | No | Yes | No | Yes |
| Director controls | No | No | No | No | No | No | No | No | No | No | No | Yes | Yes |

Robust 95% confidence intervals in parentheses, *** p<0.001, ** p<0.01, * p<0.05

**Table 3. Predicted change in probability of government committee**

| | (14) | (15) | (16) | (17) | (18) | (19) |
|---|---|---|---|---|---|---|
| Elite Network (ref=Not in largest component) | | | | | | |
| *Network core* | .084*** | .070*** | .081*** | .083*** | .048*** | .025*** |
| | (.080 - .089) | (.066 - .073) | (.077 - .085) | (.079 - .087) | (.045 - .050) | (.024 - .027) |
| *Local brokerage* | .034*** | .029*** | .035*** | .030*** | .023*** | .016*** |
| | (.032 - .035) | (.028 - .031) | (.034 - .037) | (.029 - .032) | (.022 - .024) | (.015 - .016) |
| *Largest component* | .006*** | .005*** | .006*** | .006*** | .005*** | .004*** |
| | (.006 - .006) | (.005 - .006) | (.006 - .006) | (.005 - .006) | (.005 - .005) | (.004 - .004) |
| Top rank board (ref=Top 5001-) | | | | | | |
| *Top 50* | .006*** | .008*** | .009*** | .010*** | .008*** | .009*** |
| | (.005 - .007) | (.007 - .009) | (.008 - .010) | (.009 - .011) | (.007 - .009) | (.008 - .010) |
| Top 51-500 | .005*** | .005*** | .006*** | .007*** | .005*** | .005*** |
| | (.004 - .005) | (.004 - .005) | (.005 - .006) | (.006 - .007) | (.004 - .005) | (.004 - .005) |
| Top 501-5000 | .002*** | .002*** | .002*** | .003*** | .002*** | .002*** |
| | (.002 - .002) | (.002 - .002) | (.002 - .002) | (.002 - .003) | (.001 - .002) | (.002 - .002) |
| Top 50 executive (=1) | .007*** | .012*** | .012*** | .013*** | .011*** | .006*** |
| | (.006 - .009) | (.011 - .014) | (.010 - .013) | (.011 - .015) | (.009 - .012) | (.005 - .007) |
| Top 50 chair (=1) | .004*** | .004*** | .003*** | .003*** | .005*** | .003*** |
| | (.002 - .005) | (.003 - .005) | (.002 - .004) | (.002 - .004) | (.003 - .006) | (.002 - .004) |
| Top 500 executive (=1) | -.001*** | .003*** | .002*** | .003*** | .001*** | .001*** |
| | (-.001 - -.001) | (.002 - .003) | (.002 - .003) | (.003 - .004) | (.001 - .002) | (.000 - .001) |
| Top 500 chair (=1) | .004*** | .004*** | .005*** | .003*** | .003*** | .002*** |
| | (.004 - .005) | (.004 - .005) | (.005 - .006) | (.003 - .003) | (.002 - .003) | (.002 - .003) |
| Business committee member (=1) | .035*** | | | | .020*** | .016*** |
| | (.034 - .036) | | | | (.019 - .021) | (.016 - .017) |
| Leader business association (current=1) | | .017*** | | | .019*** | .010*** |
| | | (.016 - .018) | | | (.017 - .020) | (.009 - .011) |
| Leader business association (previous=1) | | .023*** | | | .010*** | .010*** |
| | | (.022 - .024) | | | (.010 - .011) | (.009 - .010) |
| Leader labor union (current=1) | | | .012*** | | .025*** | .025*** |
| | | | (.011 - .013) | | (.023 - .027) | (.023 - .027) |
| Leader labor union (previous=1) | | | .029*** | | .014*** | .010*** |
| | | | (.027 - .030) | | (.014 - .015) | (.009 - .010) |
| Elected politician (current=1) | | | | .013*** | .013*** | .013*** |
| | | | | (.012 - .014) | (.012 - .014) | (.012 - .014) |
| Elected politician (previous=1) | | | | .012*** | .006*** | .007*** |
| | | | | (.011 - .013) | (.006 - .007) | (.007 - .008) |
| Director-month observations | 9,466,836 | 9,466,836 | 9,466,836 | 9,466,836 | 9,466,836 | 9,466,836 |
| Pseudo R2 | .177 | .190 | .171 | .184 | .249 | .298 |
| Number of boards FE | Yes | Yes | Yes | Yes | Yes | Yes |
| Company controls | No | No | No | No | No | Yes |
| Director controls | No | No | No | No | No | Yes |

Robust 95% confidence intervals in parentheses, *** p<0.001, ** p<0.01, * p<0.05